# Reinvestigation of the Yield of the Trinity Device

Eric B. Norman*, Keenan J. Thomas, Pedro V. Guillaumon
Nuclear Engineering Department
University of California
Berkeley, CA 94720 USA

**Abstract**

Previous published analyses of trinitite to determine the yield of the Trinity nuclear explosion used various estimates of the fraction of fission products that became incorporated in trinitite. Furthermore, these studies utilized nuclear data that has subsequently been revised. Utilizing a planar germanium detector, we observed gamma rays produced by the decays of both $^{137}$Cs and $^{239}$Pu from four samples of trinitite. From the observed $^{137}$Cs/$^{239}$Pu ratios of and a simple analysis method based on current nuclear data and weapons debris studies, we find an average yield of 9.1$\pm$4.5 kilotons from our four samples. While this result is substantially lower than the established total yield of 21 kilotons, it is in reasonable agreement that attributable to fissions in the $^{239}$Pu core alone.

*Corresponding author: ebnorman@berkeley.edu

## INTRODUCTION

The world's first nuclear explosion, Trinity, took place on July 16, 1945 in Alamogordo, New Mexico. The fissionable core in this device was approximately 6 kg of $^{239}$Pu that was surrounded by a ~120 kg tamper of natural uranium designed to slow the disassembly of the $^{239}$Pu during the explosion.[1,2] The total explosive yield has been established to be approximately 21 kilotons (kT) TNT equivalent from fissions in both the core and tamper[1]. The device was detonated at the top of a 100-foot-tall steel tower. Sand from the surrounding desert was swept up into the fireball, melted, mixed with device materials, and later condensed and fell to the ground. This material cooled and hardened into what is now known as trinitite.

Over the years, a number of analyses of trinitite have been performed to estimate the yield of the explosion[3-7]. To do so, one must determine the total number of fissions that occurred.

$$\mathbf{N(fissions)} = [\mathbf{N(A)}] \big/ [\mathbf{Y(A)} \mathbf{\ x\ F}] \qquad \mathbf{(1)}$$

where: N(A) = number of fission product A atoms contained in the trinitite sample

Y(A) = $^{239}$Pu fast fission yield of A

F = fraction of the produced A atoms that were incorporated in the trinitite sample

Measuring the number of a particular radioactive fission product present in a sample of trinitite is not particularly difficult. The harder problem is estimating F, the fraction of the fission product that ended up in the trinitite sample under study.

The first such yield determination was performed by Herbert Anderson of the Manhattan Project. Anderson drove a lead-lined tank into the debris field and collected the first trinitite samples. Although no details are given[2], it is stated that using "radiochemical measurements" Anderson obtained a yield of 18.4 kT. Later, such determinations produced a significant range of results[3-7]. In 1995, Atkatz and Bragg[3] used a 7.6x7.6-cm sodium iodide detector to measure gamma rays emitted from a small sample of trinitite. With its 30-year half-life, the only gamma-ray emitting fission product still observable today is $^{137}$Cs with its characteristic 662-keV line. Due to their low energies and small emission probabilities, no $^{239}$Pu gamma rays were observed. In order to estimate the fraction of all the $^{137}$Cs produced by Trinity present in their trinitite sample, these authors assumed that 1% of all the fission products ended up on the ground uniformly distributed over an area of 3.8 km$^2$. Using the area of their specimen, they estimated the $^{137}$Cs fraction in their sample. Using the cumulative yield of $^{137}$Cs from the fast neutron induced fission of $^{239}$Pu, a yield of 13 kT was inferred. In 1997, Schlauf *et al.*[4] studied four trinitite samples using a high-purity germanium detector. They too, observed $^{137}$Cs gamma rays but not those from $^{239}$Pu decay. Using the same analysis method as Atkatz and Bragg, they found yields that varied by a factor of four and averaged 40 kT.

In the years since 1997, much has been learned about the timescale for the formation of solid objects in a nuclear fireball and about the yield of $^{137}Cs$ from the fission of $^{239}Pu$. Cassata et al.[8] state " Fallout particles are formed when molten or partially molten surface materials are entrained in the hot, radioactive fireball, where they interact with condensing bomb, fission, and surface materials that were vaporized by the explosion". Furthermore, the work of Cassata *et al.*[8] and that of Lewis[9] have shown that the timescale for the formation of solid objects from low-yield devices is on the order of seconds after the explosion. This calls into question one of the key assumptions made in Refs. 3 and 4. Figure 1 illustrates the A = 137 fast fission yields from $^{239}Pu$. The volatile $^{137}I$ and $^{137}Xe$ precursors to $^{137}Cs$ would not have enough time to decay before the formation of trinitite and, due to its nature as noble gas nature, it is very unlikely that $^{137}Xe$ would be incorporated into solid materials. Thus in the analysis described below, we used the independent yield of $^{137}Cs$ of (0.69 $\pm$ 0.25)% and not the cumulative yield of (6.36 $\pm$ 0.12) % (Ref. 10).

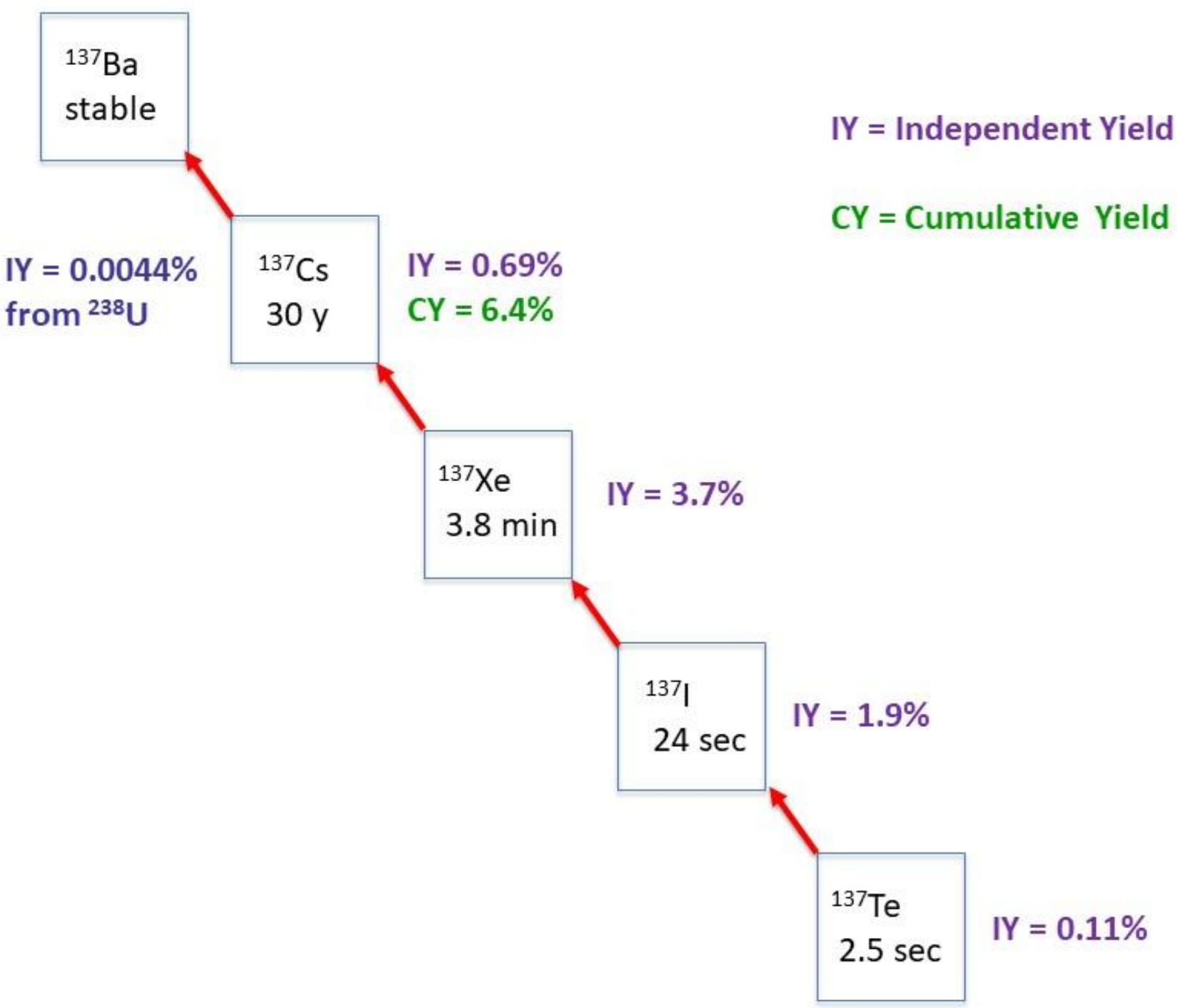

Figure 1. A = 137 fast fission yields from $^{239}Pu$.

Using aerial photos of the Trinity site, Hermes and Strickfaden[5] determined that the trinitite layer extended only 300 meters from ground zero. Wallace et al.[11], estimated this layer extended 370 meters. Taking the average of these two estimates results in the area of the trinitite deposit to be ~ 0.353 $km^2$ not the 3.8 $km^2$ assumed in Refs. 3 and 4. Somewhat remarkably, the effects of the modern values for the $^{137}Cs$ independent yield and the area of the trinitite layer nearly cancel each other out in the method of analysis used in Refs. 3 and 4. Thus the results reported in these two paper remain essentially unchanged by these developments.

## EXPERIMENT

In light of these advances in nuclear forensics, we decided to perform a new investigation of the yield of the Trinity device utilizing the same basic idea as done in Refs. 3 and 4. We purchased four trinitite samples ranging in mass from 9.4 to 16.3 grams from the Mineralogical Research Co. (MRC) in San Jose, CA. These samples are shown in Figure 2. Where these samples were collected relative to ground zero is unknown. We counted each sample for periods of 7 to 11 days using a 36-mm diameter by 13-mm thick planar germanium detector shielded with 1.27 cm of copper and 5-10 cm of lead. This detector has a thin Be window providing excellent sensitivity to low energy gamma rays and x-rays. As can be seen in Figures 3 and 4, we were able to clearly observe the 662-keV line from $^{137}$Cs decay but also the characteristic 52- and 129-keV lines from $^{239}$Pu decay. Statistical uncertainties in all of the gamma ray peak areas were $\leq$ 10%. We performed detector efficiency measurements using calibrated sources of $^{241}$Am, $^{152}$Eu, and $^{137}$Cs. In order to account for the extended nature of our trinitite samples and for gamma ray attenuation within these samples, we positioned the calibration sources at a number of different locations in front of and behind each trinitite sample. The results of these measurements were averaged to obtain an effective detection efficiency result for each sample at 52, 129, and 662 keV. Variations in the detection efficiencies between samples were $\leq$ 25%.

## RESULTS

From all of these measurements, we obtained the numbers of $^{137}$Cs and $^{239}$Pu atoms present in our samples decay corrected to July 16, 1945. These results are shown in Table I. Note that the major source of uncertainty in the number of $^{137}$Cs atoms we infer comes from the 36% uncertainty in its independent fission yield from the fast fission of $^{239}$Pu.

Wallace et al.[11] performed laser ablation-inductively coupled plasma mass spectrometry with high spatial resolution on 14 trinitite samples purchased from MRC. Their results showed that the $^{137}$Cs and $^{239}$Pu were both located primarily in the glassy component of the trinitite. They also stated that "the arid conditions of New Mexico's desert have likely prevented mobilization and leaching of long-lived radionuclides". Thus, modern studies should produce results representative of the original composition of trinitite.

Our results for $^{137}$Cs are consistent with those found in a number of previous gamma-ray studies of trinitite [4,7,12-16]. Parekh et al.[13] reported $^{137}$Cs and $^{239}$Pu activities from a study of 16 samples of trinitite that were collected in 2002 at locations of 40 – 65 meters from ground zero. These samples were analyzed together as sample A in this study. In addition, they reported $^{137}$Cs values obtained from two samples (labeled B and C) that were purchased from MRC and whose original locations relative to ground zero were unknown. Our $^{137}$Cs and $^{239}$Pu results are substantially higher than those reported by these authors from the hand-collected samples. However, our $^{137}$Cs results agree with those reported by these authors from their MRC samples. Parekh et al.[13] suggested that the differences in the results between their sample A and those from B and C could indicate that B and C came from locations further from ground zero than A.

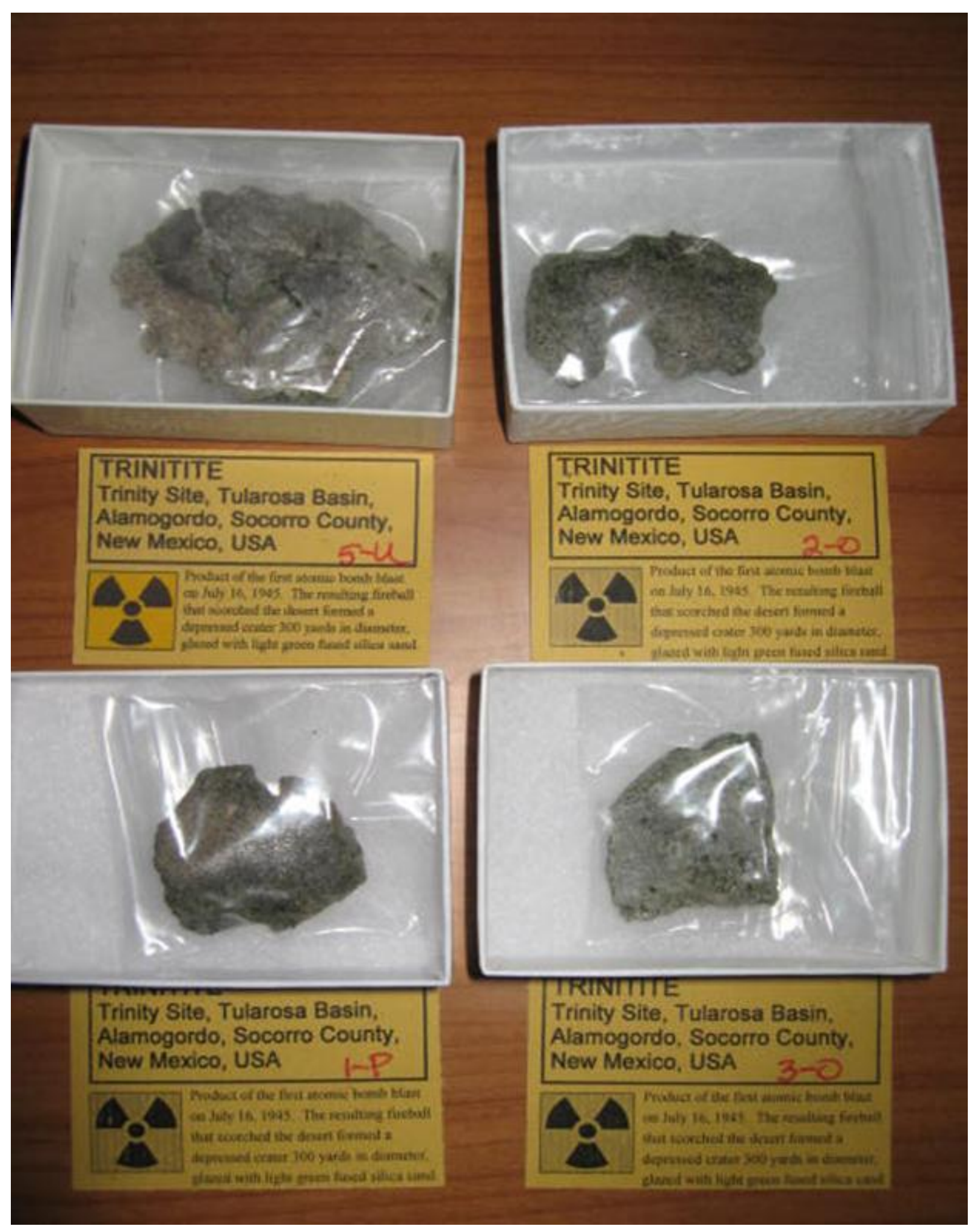

Figure 2. Trinitite samples analyzed in this investigation.

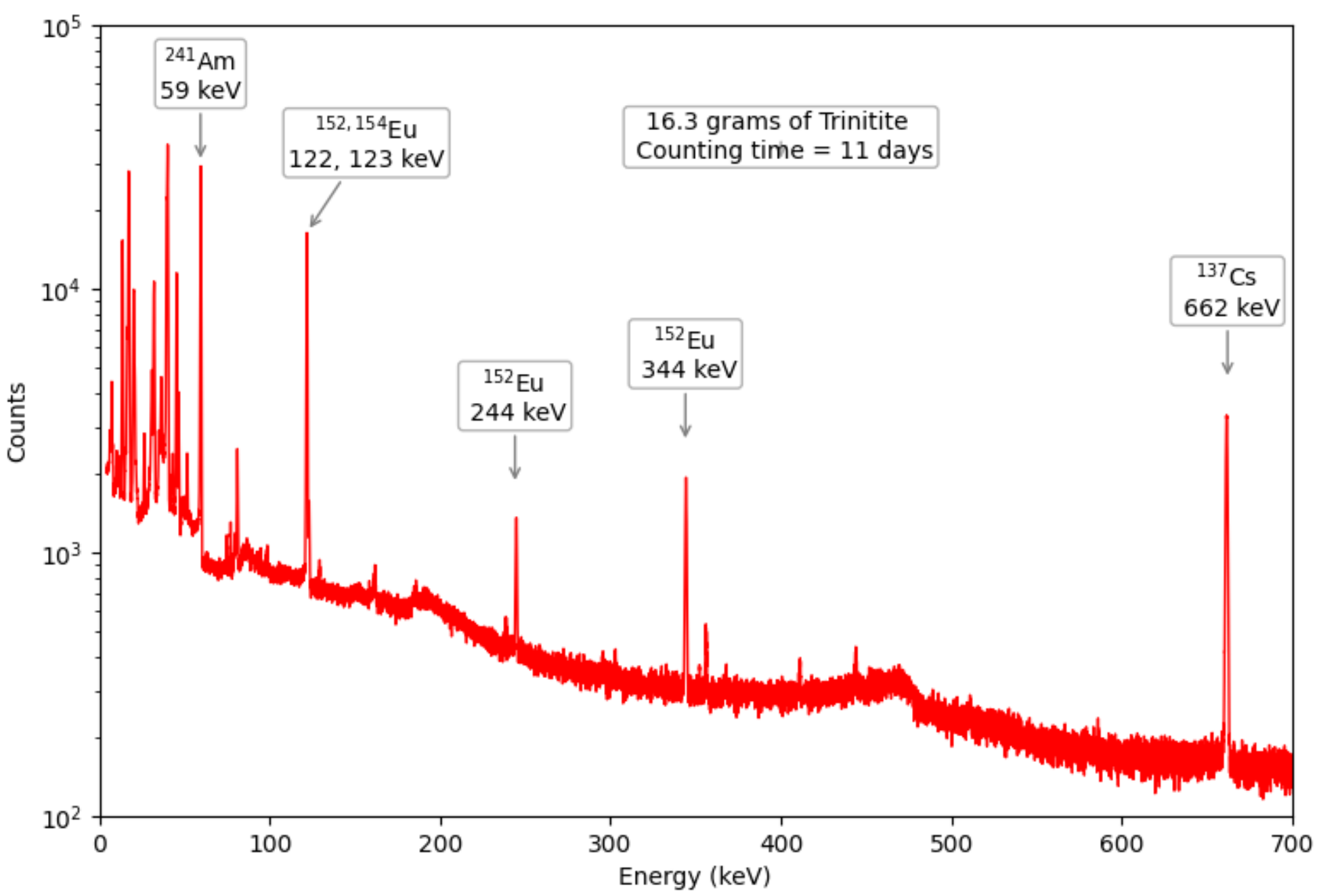

Figure 3. Gamma-ray spectrum observed from one of the trinitite samples.

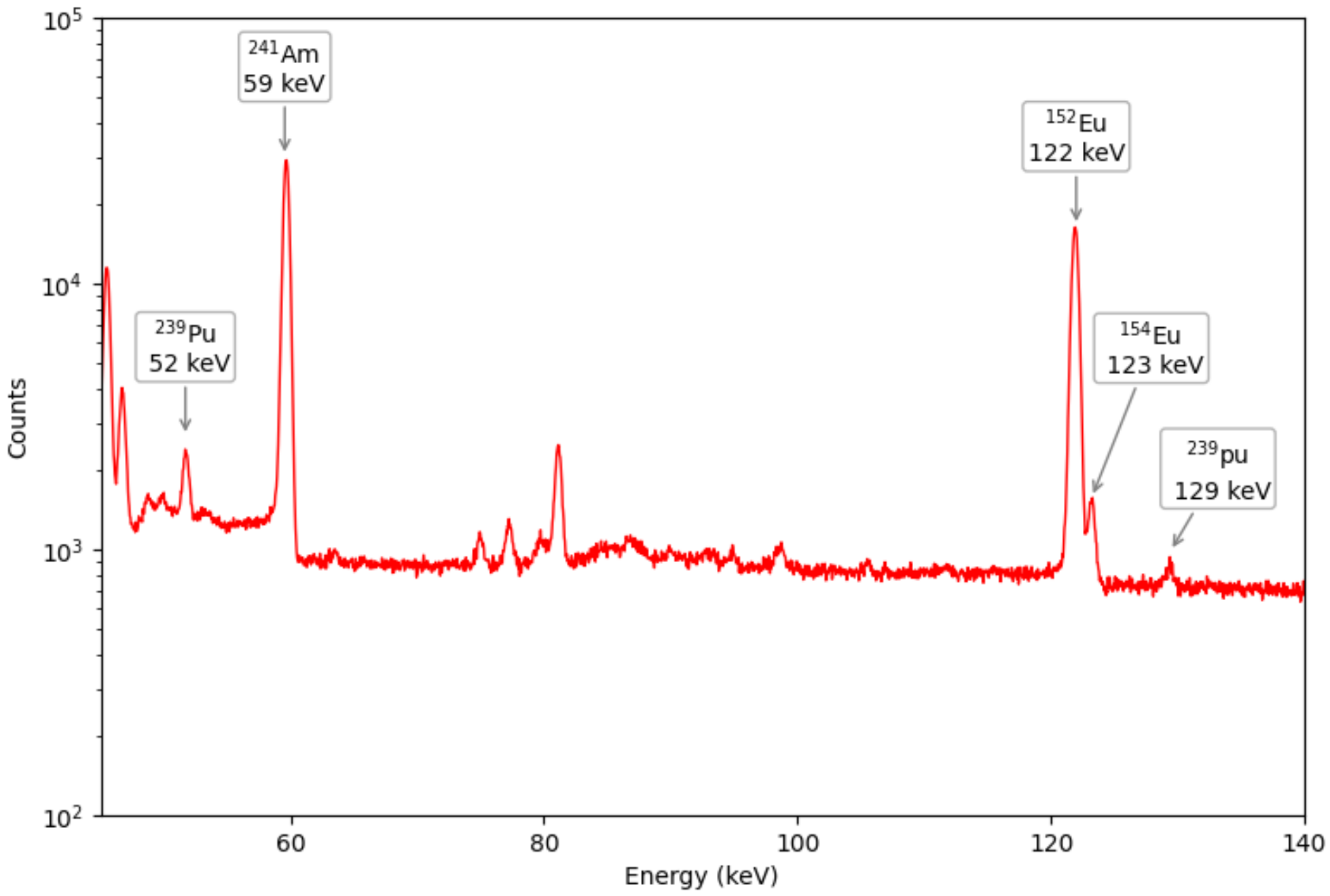

Figure 4. Low-energy portion of the gamma ray spectrum shown in Figure 3.

Although Cs is considerably more volatile than Pu, we made the simplifying assumption that the same fraction of these two elements dispersed by the explosion ended up in each of our trinitite samples. This assumption undoubtedly results in an underestimate of the yield. The fraction, of the original $^{239}$Pu that did not undergo fission, and thus was dispersed by the explosion, is given by:

$$\boldsymbol{U = 1 - \frac{1}{1 + \left(\frac{N_{Pu}}{N_{Cs}}\right) Y}} \qquad \textbf{(2)}$$

Where $N_{Pu}$ and $N_{Cs}$ are the numbers of $^{239}$Pu and $^{137}$Cs atoms originally contained in each trinitite sample. We then utilized U and the observed amounts of plutonium to infer F, the fraction of the bomb debris present in each of our trinitite samples. Thus by combining the numbers of $^{137}$Cs and $^{239}$Pu atoms in each sample we used Eqn. 1 to obtain the number of fissions that occurred. Using a value of 188 MeV/fission of $^{239}$Pu (Ref. 17), we inferred the explosive yield of the device.

| Sample | Mass (g) | N ($^{137}$Cs)/g | N ($^{239}$Pu)/g | Yield (kT) |
|---|---|---|---|---|
| 1p | 9.46 | $1.01 \times 10^{11}$ | $1.52 \times 10^{14}$ | 9.9 |
| 2o | 9.42 | $1.19 \times 10^{11}$ | $1.68 \times 10^{14}$ | 10.5 |
| 3o | 11.55 | $1.47 \times 10^{11}$ | $2.40 \times 10^{14}$ | 9.2 |
| 5u | 16.3 | $2.98 \times 10^{10}$ | $6.63 \times 10^{13}$ | 6.9 |

Average = 9.1 $\pm$ 4.5 kT

Table I. Results of our measurements of $^{137}$Cs and $^{239}$Pu from four samples of trinitite and our estimates of the yield of the Trinity explosion. The uncertainty in the average yield includes the effects of gamma ray statistics ($\leq$ 10%), detection efficiency variations ($\leq$ 25%), the 36% uncertainty in the independent fission yield of $^{137}$Cs, as well as the variation in yields determined from each sample.

Our yield results differ from each other by up to 50%, presumably due to the variable distribution of material within the mushroom cloud. Averaging our results produces a yield of 9.1 $\pm$ 4.5 kT. Note that Parekh et al.[13] did not use the $^{137}$Cs and $^{239}$Pu results from their sample A to infer an explosive yield. However, applying our analysis method to their data, we obtain a yield of 6.1 kT.

## CONCLUSIONS

As expected, our result is substantially lower than the established total yield of 21 kT. However, as pointed out by Semkov et al.[18], approximately 69% of the yield (14.5 $\pm$ 0.4 kT) is attributable to the fission of $^{239}$Pu while 31% (6.5 $\pm$ 0.4 kT) was produced by fissions of $^{235}$U and $^{238}$U contained in the tamper. The independent yields of $^{137}$Cs produced by fast fissions of $^{235}$U and

$^{238}U$ (0.122% and 0.0044%, respectively)[10] are much smaller than that from $^{239}Pu$ and thus could only produce a few percent of the $^{137}Cs$ we observed in our trinitite samples. Thus, our yield result should be considered as that due to fissions in just the $^{239}Pu$ core and is not far from the value obtained by Semkov et al[18].

From this investigation, we demonstrate that a planar germanium detector, directly measuring the gamma rays from $^{239}Pu$ and $^{137}Cs$ decays and a very simple analysis method can provide a reasonable estimate of the yield from the fissions of $^{239}Pu$ in the Trinity explosion. Our approach not only sheds new light on the historical event, but also underscores its continued significance in modern nuclear forensic techniques.

**ACKNOWLEDGEMENTS**

We wish to thank David Weiss for a careful review of this manuscript. This work was partially supported by the U. S. Department of Energy under Grant Number DE-NA0000979.